\documentclass[aps,pra,twocolumn,superscriptaddress,longbibliography,footnote]{revtex4-2}
\usepackage{bm}
\usepackage{amsmath,mathtools}
\usepackage{amssymb}
\usepackage{amsthm}
\usepackage{amsfonts}
\usepackage{graphicx}
\usepackage{epsfig}
\usepackage{epstopdf}
\usepackage[hidelinks,colorlinks=true,linkcolor=blue,citecolor=blue,anchorcolor=blue,filecolor=blue,menucolor=blue,runcolor=blue,urlcolor=blue]{hyperref}
\usepackage{color}
\usepackage{setspace}

\begin{document}

\title{Synthetic half-integer magnetic monopole and single-vortex dynamics in spherical Bose-Einstein condensates}
\author{Xi-Yu Chen}
\email{These authors contribute equally.}
\affiliation{Institute of Modern Physics and school of Physics, Northwest University, Xi'an 710127, China}
\author{Lijia Jiang}
\email{These authors contribute equally.}
\affiliation{Institute of Modern Physics and school of Physics, Northwest University, Xi'an 710127, China}
\affiliation{Shaanxi Key Laboratory for Theoretical Physics Frontiers, Xi'an 710127, China}
\affiliation{Peng Huanwu Center for Fundamental Theory, Xi'an 710127, China}
\author{Wen-Kai Bai}
\affiliation{Institute of Modern Physics and school of Physics, Northwest University, Xi'an 710127, China}
\affiliation{Shaanxi Key Laboratory for Theoretical Physics Frontiers, Xi'an 710127, China}
\affiliation{Peng Huanwu Center for Fundamental Theory, Xi'an 710127, China}
\author{Tao Yang}
\email{yangt@nwu.edu.cn}
\affiliation{Institute of Modern Physics and school of Physics, Northwest University, Xi'an 710127, China}
\affiliation{Shaanxi Key Laboratory for Theoretical Physics Frontiers, Xi'an 710127, China}
\affiliation{Peng Huanwu Center for Fundamental Theory, Xi'an 710127, China}
\author{Jun-Hui Zheng}
\email{junhui.zheng@nwu.edu.cn}
\affiliation{Institute of Modern Physics and school of Physics, Northwest University, Xi'an 710127, China}
\affiliation{Shaanxi Key Laboratory for Theoretical Physics Frontiers, Xi'an 710127, China}
\affiliation{Peng Huanwu Center for Fundamental Theory, Xi'an 710127, China}

\begin{abstract}
{Magnetic} monopoles are crucial in explaining the quantization of electric charges and quantum hall effects, while artificially creating a minimal {magnetic} monopole in experiments remains a challenge. Here, we come up with a flexible way to simulate a half-integer-type monopole in Bose gases and investigate the induced vortex dynamics on a sphere. We list the possible experiment parameter settings for different isotopes and discuss their experimental feasibility. With the assumption of a rigid monopole-vortex structure, we analytically predict the vortex trajectory in an external magnetic field. We then confirm the result by numerically solving the Gross-Pitaevskii equation, which employs two gauges {simultaneously (the Wu-Yang approach)} to prevent singularity in the one-gauge method when a monopole is present. The study offers significant insight into the characteristics of monopoles and vortices, facilitating avenues for experimental validation.
\end{abstract}
\maketitle

\section{Introduction}

{Magnetic} monopoles, which are particles with isolated magnetic charges, have been a subject of speculation in theoretical physics. Although the existence of elementary magnetic monopoles remains a mystery \cite{Dirac1931,Yang1975,Yang1976,Drukier1979,Friseh1990,Shellard2000,Guth1981}, analogues emerge in condensed matter systems, for instance, on the interfaces of rotating superfluid mixtures \cite{Salomaa1987}, at the ends of spin chains in spin ices \cite{Castelnovo2008,Gingras2009,Bramwell2009,Giblin2011} and that of skyrmion-line excitations in chiral magnets \cite{Lin2016}, or in the momentum space of quantum Hall systems \cite{Fang2003}. Monopoles can also be {mimicked by} creating a thin magnetic needle \cite{Boxem2013}, the Berry curvature of a varying superconducting qubit state \cite{Zhang2017}, or through magneto-optical effects \cite{Edelstein2024}.

\begin{figure}
  \centering
  \includegraphics[width=\columnwidth]{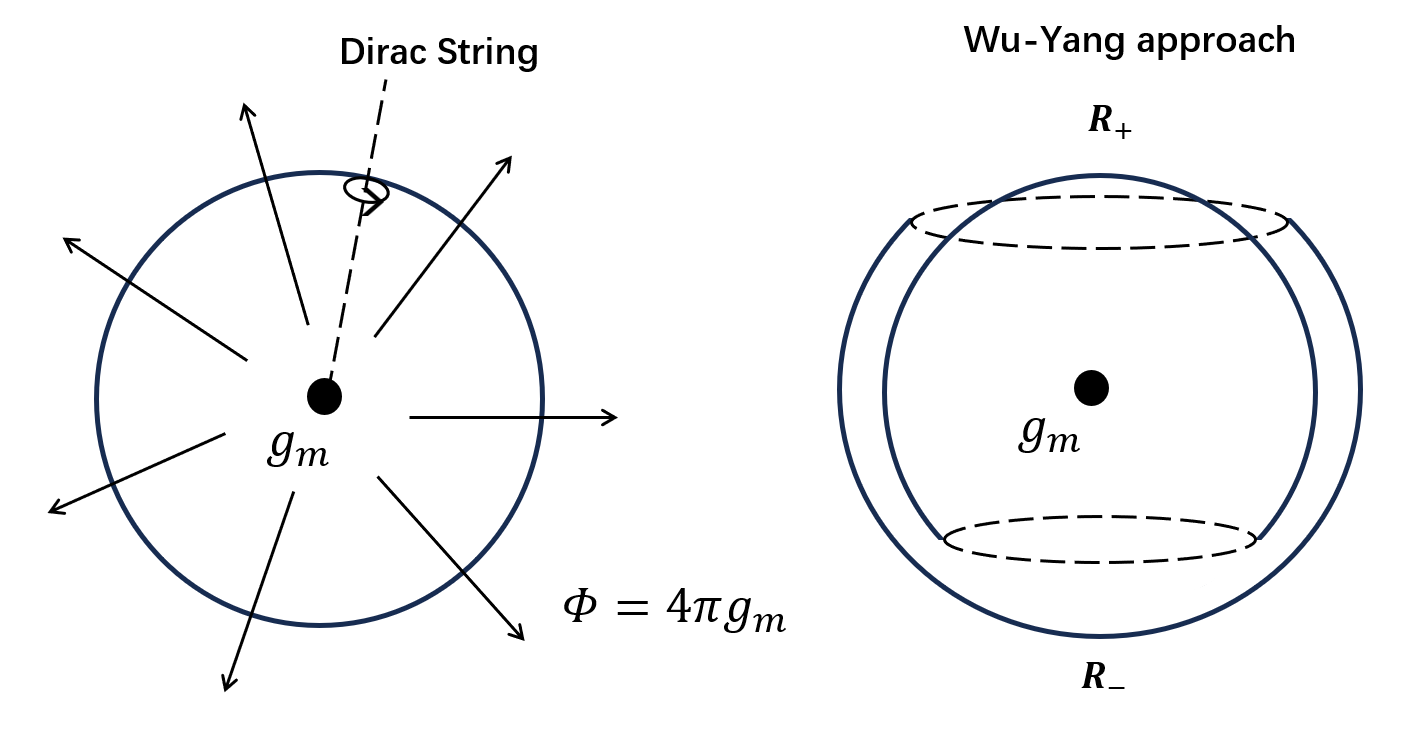}\\
  \caption{  Dirac string and the Wu-Yang approach. When the particle moves around the string closely, the Aharonov-Bohm phase is $e^{i q_e \Phi}$, where $\Phi = 4\pi g_m$ is the magnetic flux. In the Wu-Yang approach, two different gauges are needed to avoid the Dirac string.
}\label{fig0}
\end{figure}

The existence of magnetic monopoles implies that the magnetic field is no longer sourceless, i.e., $\nabla \cdot \bm B \neq 0$. To retain the electromagnetic vector potential $\bm{A}(\bm r)$ in this scenario, {it is necessary to introduce a singular line in $\bm{A}(\bm r)$, known as the Dirac string.} The Dirac string shows its physical effects, when a charged particle (with electric charge $q_e$) moves around the string closely. The particle acquires an Aharonov-Bohm phase $e^{i q_e \Phi}$, where $\Phi = 4\pi g_m$ is the magnetic flux ($g_m$ is the magnetic charge) passing through a surface (which does not intersect the Dirac string) bounded by the particle's trajectory, as illustrated in Fig.\ref{fig0}. {Crucially, since the Dirac string is hypothetical and unobservable, it must yield no detectable consequences. This requires the phase factor to satisfy  $q_e \Phi = 2 n \pi $ with $n \in \mathbb{Z}$ (in natural units), leading directly to the Dirac charge quantization condition: $2g_m q_e \in \mathbb{Z}$  \cite{Dirac1931,Yang1975}.} This duality explains why electric charges are quantized in the universe if a magnetic monopole exists, i.e. $q_e = n/ 2g_m$. Conversely, it also restricts magnetic charges to be integer multiples of $1/2 q_e$.

In experiments of cold atom simulation with Bose-Einstein condensates (BEC), the effective magnetic field originates from the Berry curvature, which stems from the adiabatic evolution of the atomic spin state  $|\chi (\bm r)\rangle$ in space. The corresponding Berry connection, $\bm A(\bm r) = \langle \chi(\bm r) | i\nabla |\chi(\bm r) \rangle$, serves as an effective gauge field, which introduces a geometric phase $e^{i\int \bm A(\bm r) \cdot d \bm r}$ into the atomic spatial wavefunction, depending on the path. {By analogy with the form of the Aharonov-Bohm effect, the effective charge in the geometric phase is identified to $q_e=1$.} Synthetic magnetic monopoles can be constructed by aligning the spin of the atoms (with the quantum number $F$) with a special configuration in space so that moving atoms experience a Berry curvature corresponding to a monopole with $g_m =F$ \cite{Ray2014}. From the quantization condition, the allowed magnetic charge is either an integer or a half-integer. However, magnetic monopole realized in the BEC simulations so far, contains an integer magnetic charge \cite{Ray2014,Pietil2009,Ray2015,Ruokokoski2011} since the atomic spin quantum number $F$ must be an integer for bosons. Whether and how half-integer-type magnetic monopoles, especially the minimal monopole, can be realized in boson gases remains an open question.

With a magnetic monopole situated in the center of a sphere, a uniform magnetic field perpendicular to the spherical surface is created. For the noninteracting case, the single-particle wave functions on the sphere surface, the so-called monopole harmonics, have been analytically obtained by Wu and Yang \cite{Yang1976}. The eigenstate has vortices with a total vortex charge of $2 g_m$. Especially for the $g_m=1$ bosonic case, the ground state contains two vortices \cite{Yang1976} and these vortices can {be manipulated to desired position} \cite{Xu2024}. In the interacting case, for a BEC system with an integer $g_m$, total $2g_m$ vortices exist in the ground state and form stable lattice patterns, like Thomson plum pudding model \cite{Zhou2018}. So far, the vortex dynamics on the spherical BEC in the literature mainly focus on cases in the absence of magnetic monopoles, such as vortex-antivortex dynamics, BKT phase transition, and the formation of vortex clusters \cite{Turner2010,Tononi2022,Guenther2020,Kanai2021,Yu2024,Dritschel2015,Sun2020,Bereta2021,Caracanhas2022,Efimkin2023,White2024,He2023}. Yet, it is unclear how a monopole influences the dynamics of vortex.

On the other hand, the spherical system with a monopole serves as an ideal framework for investigating fractional quantum hall effects \cite{Haldane1983} and vortex dynamics without boundary effects in a finite size \cite{Kanai2021}. It is imperative to examine the realization of the minimal monopole and the characteristics of the minimal vortex-monopole unit in the spherical system, which underpins the study of intricate vortex structures in the presence of monopoles with high magnetic charges. In this article, we explore the realization of a magnetic monopole with a half-integer magnetic charge in a cold atom system, as well as the dynamics of a single vortex (in the presence of a minimal monopole) driven by an external field. The rest of the article is organized as follows:

In Sec.\,\ref{sec2}, we propose a flexible scheme to realize a magnetic monopole with a half-integer magnetic charge in the alkali- or hydrogen-atom system. The key steps are to freeze the electron's angular momentum $(\bm J)$ and polarize the nuclear spin (with a half-integer quantum number $I$) aligning to the direction of the atom's position vector, leading to an effective radial magnetic field corresponding to a monopole field of $g_m = I$.

In Sec.\,\ref{sec3}, we develop the ground state of spherical BEC in the presence of a monopole with a minimal charge $g_m =1/2$. For strong interaction, we take the trial wave function $\psi(\theta,\phi) \propto \tanh(b\theta) e^{i\phi}$ for the ground state with a vortex locating at the north pole (in spherical coordinates). The fitting function agrees very well with the results obtained from the numerical calculation. We generalize the solution by rotating the vortex core to an arbitrary position. We demonstrate that each atom in this vortex-monopole system possesses angular momentum of magnitude ${\hbar}/{2}$ orientating to the vortex position.

In Sec.\,\ref{sec5}, we study the vortex dynamics of spherical BEC with a monopole $g_m =1/2$ and an additional external electromagnetic field. Assuming a rigid vortex-monopole configuration, we demonstrate that the vortex-monopole structure performs a precession motion and develop an equation of motion to describe the trajectory of the vortex center.

In Sec.\,\ref{sec4}, we employ the Gross-Pitaevskii equation (GPE) to simulate the vortex dynamics in a 3D shell-shaped system with a magnetic monopole, where a `bi-gauge evolution' approach is developed to avoid singularities of the gauge fields in the numerical calculation. The numerical results are consistent with our theoretical prediction.

Our work offers a scheme for the realization of magnetic monopoles with half-integer charges, validates the point-vortex picture for the strong interaction in the presence of monopoles, and advances the study of the dynamics of the simplest vortex-monople structure.

\section{Realizing monopoles of half-integer magnetic charges}\label{sec2}
In this section, we present the scheme of realizing monopoles with half-integer magnetic charges. Considering an atom in the presence of a magnetic field $\bm B$, the Hamiltonian describing the energy level splitting for the internal degree of freedom of atoms contains the hyperfine interaction and Zeeman energies,
\begin{equation}\label{eq1}
\hat{H}_{\text{spin}}=A\bm{\hat I}\cdot\bm{ \hat J}+g\mu_{B}\bm{B}\cdot\bm{\hat J}-(\mu_I/I) \bm{B}\cdot\bm{\hat I},
\end{equation}
where $A$ is a constant, {$\bm{\hat I}$ is the nuclear spin, $\bm{\hat J}$ denotes the total angular momentum of electron in the atom}, $g\simeq 2$ is the Land\'e $g$-factor, $\mu_{B}$ is the Bohr magneton, and $\mu_I \bm{\hat I}/I$ is the nuclear magnetic moment \cite{Pethick2008}. {The total angular momentum of electron is $\bm{\hat J} = \bm{\hat L} + \bm{\hat S}$, where $\bm{\hat L}$ is the electron's orbital angular momentum and $\bm{\hat S}$ is the electron spin. The atomic spin is defined as $\bm{\hat{F}}= \bm{\hat{I}} + \bm{\hat{J}}$}. For a ground-state alkali or hydrogen atom, the electrons have no orbital angular momentum ($L=0$), so the electron's total angular momentum quantum number is equal to its spin quantum number, that is $J = S = 1/2$ \cite{Pethick2008}. In this scenario, if the atom is a boson, the nuclear spin must be half-integer, resulting in an integer atomic spin quantum number $F$. To realize the half-integer type of magnetic monopoles in these bosonic systems, we adopt the following two steps:

First, a strong magnetic field $\bm{B}_0=B_{0}{\bm e}_z$ with a strength of $B_0\gg A I/g \mu_{B}$ is applied to the system, so that the second term in Eq.\eqref{eq1} is much larger than the hyperfine interaction. {Note that the nuclear magnetic moment is three orders of magnitude smaller than the electron magnetic moment}. Thus, the second term of $\hat{H}_\text{spin}$ is dominant, and the angular momentum $\bm J$ is locked to the state $|J, -J\rangle_{{z}}$.  In this manifold, the second term becomes $- g\mu_{B} B_0 J$, { and $A {\hat I_z}{\hat J_z}$ from the first term becomes $A \hat I_z/2$ (here, $J=1/2$ has been used). The other terms $A \hat I_x  \hat J_x + A \hat I_y  \hat J_y  = A (\hat I_+ \hat J_-+ \hat I_- \hat J_+)/2$ couples the high energy manifold $|J, J\rangle_{{z}}$, which will contribute a second order correction.} Together with the last term, $-(\mu_I/I) B_0 {\hat I_z}$, the low-energy effective Hamiltonian in the low-energy manifold $|J, -J\rangle_{{z}}$, up to the second order approximation, becomes
\begin{eqnarray}\label{Hspin}
 \hat{H}_{\text{spin}} &\simeq& -\omega_L \hat{I}_z - \frac{A^2}{4 g \mu_B B_0} {_{z}\langle J, -J | \hat J_-\hat I_+ \hat I_- \hat J_+ |J, -J\rangle_{{z}}} \notag\\
 && - g\mu_{B} B_0 J ,
\end{eqnarray}
where $\omega_L = A/2 +\mu_I B_{0}/I$. Using the fact that $J =1/2$, we have
\begin{equation}
{_{z}\langle J, -J | \hat J_-\hat I_+ \hat I_- \hat J_+ |J, -J\rangle_{{z}}} = I^2 -\hat I_z^2 + \hat I_z,
\end{equation}
Considering the condition $B_0\gg A I/g \mu_{B}$ and after discarding a constant term, we approximately have
\begin{equation}\label{hspinx}
\hat{H}_{\text{spin}} \simeq -\omega_L \hat{I}_z,
\end{equation}
which represents the energy splitting for the nuclear spin. This is different from the weak magnetic field case, where the low-energy physics is captured by the Zeeman splitting of the total spin $\bm F$ \cite{Pietil2009}. Note that we have neglected the second term (the second order correction) of Eq.\,\eqref{Hspin} for simplicity, which contributes {a linear $\propto\hat I_z $} and a quadratic term $\propto\hat I_z ^2$. {The linear term slightly corrects the magnitude of $\omega_L$ and does not change the form \eqref{hspinx}}. The quadratic term can be compensated by the Floquet mechanics as shown below. For  $\bm{I}=1/2$, which is our main focus for generating minimal monopoles, it is always a two-level system and the quadratic term becomes a constant term, so the second term of Eq.\,\eqref{Hspin} just contributes a correction to the splitting $\omega_L$.

Second, similar to the proposal in Ref.\,\cite{Zhou2018}, a periodic oscillating {quadrupolar} field
\begin{equation}
\bm{B}_1= B'_{1}(1-4\lambda_0\cos{\omega t})(x \bm e_{x}+y\bm e_{y}-2z \bm e_{z}),
\end{equation}
is applied. The quadratic term mentioned in above can be {compensated by a proper choice of $\lambda_0$} \cite{Zhou2018}. Here we focus on $\lambda_0 =1$. We require $\omega = \omega_L$ to couple different nuclear spin states resonantly. In the rotating frame of
\begin{equation}
    \hat{U}=\exp(i\omega_{L}t\hat{I}_{z}),
\end{equation}
the Hamiltonian becomes
\begin{equation}
\hat{H}_{I}=\hat{U}^{\dagger}\big[ -\omega_L \hat{I}_z -(\mu_I/I) \bm{B}_1\cdot\bm{\hat{I}}\big]\hat{U}-i\hat{U}^{\dagger}\partial_{t}\hat{U}.
\end{equation}
The low-energy effective Hamiltonian in the rotating wave approximation \cite{Zhou2018}, where $\omega_L \gg \mu_I B'_1 L/I$ being much larger than all other energy scales ($L$ is the typical size of the BEC {and $B'_1 L$ charactorizes the magnitude of the quadrupolar magnetic field}),  becomes
\begin{equation}
\hat{H}_{I}= 2\frac{\mu_I}{I}B'_{1}\bm{\hat{I}}\cdot\bm{{r}}.
\end{equation}

{When} the two magnetic fields $\bm B_0$ and $\bm B_1$  are applied, the lowest spin state reads
\begin{equation}\label{lowest}
    |\chi(\bm r)\rangle_{\pm} =|J, -J\rangle_z \otimes |I, -I\rangle_{{r}}^{\pm},
\end{equation}
where
\begin{equation}
    |I, -I\rangle_{{r}}^{\pm}= e^{\mp i I \phi} \exp(-i\hat{I}_{z}\phi) \exp(-i\hat{I}_{y}\theta)|I, -I\rangle_{z}
\end{equation}
is the common eigenstate of $\bm{\hat{I}}^2$ and $\bm{\hat{I}} \cdot \bm e_{r}$. Here, $(\theta, \phi)$ are the polar and azimuthal angles, respectively. In contrast to the integer $I$ case, for a half-integer $I$, the phase $e^{\pm i I \phi}$ is necessary to ensure $|I, -I\rangle_{{r}}^{\pm}$ remains the same value after the transformation $\phi\rightarrow\phi+2\pi$, where $\pm$ labels different gauge choices \cite{Zhou2018}.

{Now we consider the full
single-particle Hamiltonian}, which includes $H_{\text{spin}}$, the external trap potential ${V}_H(r)$, and the kinetic energy. It reads
\begin{equation}
\hat{H}=-\frac{1}{2m}\nabla^{2}+{V}_H(r)+\hat{H}_{\text{spin}},
\end{equation}
where the external trap potential is ${V}_H(r)=m\omega^2_{H}\bm{{r}}^{2}/2$. Both magnetic fields, $\bm B_0$ and $\bm B_1$, are applied. In the adiabatic approximation of the lowest spin manifold \eqref{lowest}, the effective Hamiltonian governing the dynamics of atoms in space is
\begin{equation}\label{eq12}
\hat{H}_{\pm}=(-i\nabla - \bm{{A}}_\pm)^{2}/2m+{V}(r),
\end{equation}
where the vector potential is
\begin{equation}\label{eq4}
\bm{A}_{\pm}(\bm r)={_{\pm}\langle}\chi(\bm r)| {i \nabla} |\chi(\bm r)\rangle_{\pm} =\frac{(\pm 1-\cos{\theta})I }{r\sin{\theta}}\bm {e}_\phi,
\end{equation}
and the scalar potential ${V}(r) = {V}_H(r) - 2\mu_I B'_{1}r + I/2m r^2$ makes the BEC shell-shaped \cite{Zhou2018}. The reorientation of the spin state during the motion of the atom induces a gauge field and corrects the scalar potential. The effective magnetic field is $\bm B = \nabla \times \bm A = I {\bm e}_r/ r^2$. This field exactly corresponds to an effective magnetic charge of $g_m = I$, which is half-integer in our systems.

In summary, to freeze the degree of freedom of the angular momentum of electrons, it requires a strong field $B_0 \gg B_c \equiv A I/ g\mu_B$, and to employ the rotating wave approximation, it requires the {\color{red}} drive frequency to be much larger than other energy scales of the {quadrupolar} field, $\omega_L = A/2 +\mu_I B_{0}/I \gg \omega_c \equiv \mu_I B'_1 L/I$.  On the other hand, to use the adiabatic approximation for the lowest spin manifold, it is necessary to require the system's temperature is much smaller than the energy splitting $\hbar \omega_c$. These three conditions are the key to realize a half-integer type of magnetic monopoles. In Table \ref{table1}, we list the value of $B_c$ using the known experimental data, estimate the value of $\omega_c$ by assuming that the typical size of the BEC is $L = 100~\mu$m and the magnetic field gradient is $B'_1 = 0.5 ~\text{T/m}$, and evaluate $\omega_L$ by setting $B_0 = 1 ~\text{T}$, for different isotopes with a nuclear spin $I=1/2$, $3/2$, $5/2$, and $7/2$, respectively. The table shows that in this set, both conditions $B_0\gg B_c$ and $\omega_L \gg \omega_c$ are satisfied. Indeed, $\omega_L$ is of the order of $A/2$, which is always much larger than $\omega_c$. So $B_0\gg B_c$ is the only factor that should be taken into account. For hydrogen, $\hbar \omega_c \simeq 100~\text{nK}$, a uniform field with $B_0 = 0.25 ~\text{T}$ (about ten times of $B_c$) is sufficient to induce a minimal monopole of $g_m=1/2$. For $^{39}\text{K}$, $^{41}\text{K}$,$^{85}\text{Rb}$,$^{133}\text{Cs}$, $\hbar \omega_c$ is very small. Therefore, the required temperature is too low to be achievable in the current setting. A strategy to overcome this issue is to increase the radius of the shell-shaped BEC so that the energy splitting $\hbar \omega_c$ can be enhanced. Another concern is about gravitational effects, which may distort the shape of the BEC shell. {A strategy to weaken this effect involves using optical trapping to counteract the gravitational potential or employing a microgravity environment.}

\begin{table}[t]\label{table1}
\centering
  \begin{tabular}{ccccccccc}
  \hline
  \hline
  Isotope    & $I$ & ${A}$[MHz] & ${\mu_I}/{\mu_N}$ &   $B_c$[mT] & $\omega_c$[kHz] & $\omega_L$[MHz]\\
  \hline
  {$^1$H}      & 1/2  & 8925 & 2.793  & 25.4 & 13.38 & 4730 \\
  {$^7$Li}     & 3/2  & 2524 & 3.256  & 21.5 & 5.20 & 1366 \\
  {$^{23}$Na}  & 3/2  & 5566 & 2.218  & 47.5 & 3.54 & 2854 \\
  {$^{39}$K}   & 3/2  & 1451 & 0.391  & 12.4 & 0.62 & 738\\
  {$^{41}$K}   & 3/2  & 798 & 0.215  & 6.8  & 0.34 & 406\\
  {$^{87}$Rb}  & 3/2  & 21472 & 2.751  & 183.1 & 4.39 & 10824
  \\
  {$^{85}$Rb}  & 5/2  & 6358 & 1.353  & 90.4 & 1.30 & 3205\\
 {$^{133}$Cs}  & 7/2  & 14440 & 2.579  & 287.4 & 1.76 & 7255 \\
  \hline
  \hline
\end{tabular}
\caption{The parameter estimation of realizing half-integer magnetic monopoles for different isotopes. The data for $I$, $A$, and $\mu_I/\mu_N$ are selected from Refs.\cite{Diermaier2017, Arimondo1977,Pethick2008}, where $\mu_N$ is the nuclear magneton. $B_c \equiv A I/g\mu_{B}$ and $\omega_c \equiv \mu_I B'_1 L/I$ are evaluated accordingly, assuming the typical size of BEC to be $L = 100~\mu$m and $B'_1 = 0.5 ~\text{T/m}$\cite{Pietil2009,Ray2015}. $\omega_L = A/2 +\mu_I B_{0}/I$ is obtained by setting $B_0 = 1$ T. To realize monopoles, it requires $B_0\gg B_c$, $\omega_L \gg \omega_c $, and $\hbar \omega_c \gg k_B T$. }\label{table1}
\end{table}

\section{Single vortex on a sphere}\label{sec3}
In this section, we calculate the ground state of the spherical BEC in the presence of a minimal monopole with $I=1/2$. The dynamical behaviors of the BEC are described by the GPE,
\begin{equation}\label{gpe}
 i \partial_t{\psi_{\pm }}= \hat{H}_{0} \psi_{\pm},
\end{equation}
where $\hat{H}_0 = \hat{H}_{\pm} + \lambda_{3D}|\psi_{\pm}|^2$ contains the mean-field contribution from the {contact} atom-atom interaction, and the wavefunction is normalized to one in the whole space. To avoid Dirac strings---the singularities---from $\bm A_\pm$, the two gauges are utilized for $\psi_\pm$ in the regions
\begin{eqnarray}
 R_+ &\equiv& \{(\theta,\phi) | 0\leq \theta \leq \pi/2 + \delta, 0\leq\phi <2\pi\}, \\
 R_- &\equiv& \{(\theta,\phi) | \pi/2 - \delta \leq \theta \leq \pi, 0\leq\phi <2\pi\},
\end{eqnarray}
respectively, where $0<\delta<\frac{\pi}{2}$ \cite{Yang1976}. We sketch this Wu-Yang approach in Fig.\,\ref{fig0}. In the overlap region, $R_+ \cap R_-$, the wave functions are connected by the $U(1)$ gauge transformation
\begin{equation}
    S =e^{2 i I \phi}:~~ \psi_{+}=S\psi_{-},
\end{equation}
and accordingly, ${\bm A}_+={\bm A}_- - i\nabla \ln{S}$ \cite{Yang1975}. Note the density distribution is gauge invariant, {i.e., $ |\psi_{+}|^2 = |\psi_{-}|^2$}.

We calculate the ground-state wave function when {$ l_0 \equiv \mu_I B'_1/m\omega_H^2 \gg 1/\sqrt{m\omega_H}$}, in which case the BEC is narrow shell-shaped with radius $r_0 = 2 l_0$ {and thickness $\sim 1/\sqrt{m\omega_H}$} \cite{Zhou2018}. For $\lambda_{3D}=0$, the ground state of $I=1/2$ can be constructed from the monopole harmonics \cite{Yang1976}, i.e.,
\begin{equation}\label{eq8}
\psi_+(\theta,\phi; \theta_v,\phi_v) = \frac{1}{\sqrt{\pi}}\big[ e^{i(\phi-\phi_v)} \sin\frac{\theta}{2}\cos\frac{\theta_v} {2}- \cos\frac{\theta} {2}\sin\frac{\theta_v}{2}\big],
\end{equation}
which vanishes at the vortex's core position $(\theta_v, \phi_v)$ and its eigen-energy is $1/4 m r_0^2$. On the other sector, $\psi_-$ is obtained via the gauge transformation: $\psi_-= S^{-1}\psi_+$.

In a closed manifold with monopoles inside, a $2\pi$ phase change of the wave function along a loop on the manifold does not always indicate a vortex, since it may be eliminated by a gauge transformation. {Note that} the velocity field is gauge invariant. Specifically for the case $\theta_v =\phi_v =0$, we have
\begin{equation}\label{vel}
    \bm{v} = \frac{1}{m} [\nabla {\arg \psi} - \bm A] = \frac{1}{m}\frac{(1+\cos{\theta})}{2 r_0 \sin{\theta}}\bm {e}_\phi.
\end{equation}

The singularity of $\nabla\times \bm v $ indicates the position of the vortex. The winding number (i.e., the vortex charge) is given by the limit of the loop integral surrounding the vortex: $ \lim_{l \rightarrow 0} ({m}/{2\pi \hbar})\oint_l \bm v \cdot d\bm r = 1$. Beyond the vortex center, the vorticity $\nabla\times \bm v = -\bm B/m$ is nonzero everywhere due to the presence of a monopole. {The velocity field $\bm v$ cannot be expressed as the gradient of a scalar function with singular points, unlike in a system without monopoles \cite{Turner2010}.}

Note that the system is symmetric under rotations with generators $\bm{\hat{L}} = \bm{r} \times \bm{\hat p} - I {\bm{r}}/r$, where $\bm{\hat p} = -i\nabla - \bm{{A}}_\pm$ \cite{Yang1976}. The appearance of vortices breaks the rotational symmetry spontaneously.

For interacting cases, the ground state with a vortex locating at  $\theta_v =\phi_v =0$, in general, has the form
\begin{equation} \label{psig}
   \psi_+(\theta,\phi;\theta_v =0,\phi_v =0) = \frac{1}{r_0} f(\theta)  e^{i\phi} e^{-i\mu t/ m r_0^2}
\end{equation}
and $\psi_-= S^{-1}\psi_+$. We will also use {Dirac}
notation $\psi_{\pm}(\theta,\phi; \theta_v,\phi_v)\equiv \langle \theta,\phi| \psi_{\pm}(\theta_v,\phi_v) \rangle$. The velocity field is the same as Eq.\,\eqref{vel}, which is completely independent of the interaction strength. From GPE, we find that the amplitude $f(\theta)$ satisfies the gauge independent equation
\begin{equation}\label{eom}
    \mu f =  \frac{1}{2} \Big[ -\partial_\theta^2 - \cot\theta \partial_\theta   + \frac{1+\cos\theta}{4(1-\cos\theta) } \Big] f + \lambda |f|^2 f,
\end{equation}
where $\lambda = m \lambda_{2D}$ is dimensionless, $\lambda_{2D}$ is the effective two-dimensional interaction strength on the sphere (see Appendix \ref{inter}), and the normalization condition is $ 2\pi \int |f|^2 \sin\theta d\theta =1$.

\begin{figure}[t]
\includegraphics[width=0.95\columnwidth]{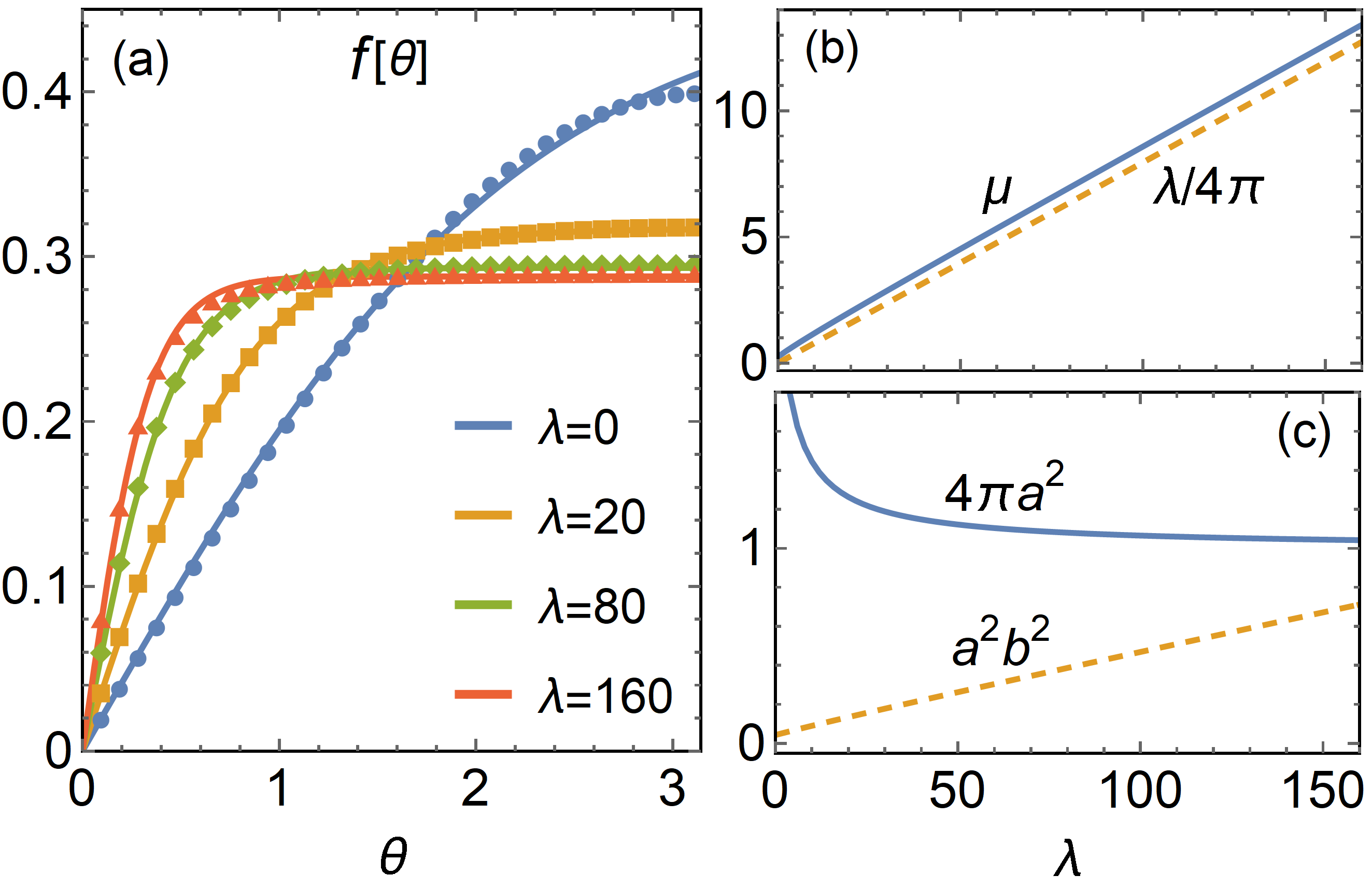}
\caption{(a) The ground-state wave functions for different inter-atom interaction strengths ($\lambda$) [symbols for the numerical results of Eq.\,\eqref{eom} and solid lines for the fitting results of the trial function]; (b) the chemical potential $\mu$ and (c) the   fitting parameters as functions of $\lambda$. For large $\lambda$, we have $b^2 \simeq 0.051 \lambda + 0.76$.}
\label{fig1t}
\end{figure}

Approximately, we employ a trial solution $f(\theta)= a \tanh(b \theta)$ with fitting parameters $a$ and $b$. The fitting results agree very well with the numerical results of Eq.\,\eqref{eom} especially for strong interaction cases, as shown in Fig.\,\ref{fig1t}. For a small $\theta$, the square of the slope of $f(\theta)$, i.e., $a^2 b^2$, linearly increases with $\lambda$, like the vortex solution in the plane case \cite{Pethick2008}. For large interactions, the density away from the vortex center becomes uniform and the vortex size is small. Consequently, $a^2 \simeq 1/ 4\pi$ and $\mu \simeq \lambda/4\pi$ (in the Thomas-Fermi approximation). The numerical result of $\mu$ [see Fig.\,\ref{fig1t}(b)] is a bit larger because the velocity fields are finite. For large $\lambda$, from Fig.\,\ref{fig1t}(c), we have $b^2 \simeq 0.051 \lambda + 0.76$ from the numerical fitting by replacing $a^2$ with $ 1/ 4\pi$. Therefore, the healing length  $\theta_c = 1/b$ decreases with the interaction strength. Correspondingly, the total particle number missing in the vortex decreases as well, and for $\theta_c \ll 1 $, we have  $n_{\text{miss}} \equiv 2\pi\int (a^2-[f(\theta)]^2) \sin\theta d\theta \simeq 0.347\theta_c^2 $.

The solution with the vortex core at an arbitrary position $(\theta_v, \phi_v)$ is obtained by the rotation transformation,
\begin{equation}\label{sol}
\psi_+(\theta,\phi;\theta_v,\phi_v) = \exp(-i\hat{L}_{z}\phi_v) \exp(-i\hat{L}_{y}\theta_v)\psi_{+}(\theta,\phi;0,0).
\end{equation}
To engineer vortex at the target position in experiments, a small positive Gaussian potential with the center at $\bm{r}_v = r_0 {\bm n}_v$ can be employed, where $ {\bm n}_v =(\sin\theta_v\cos\phi_v,\sin\theta_v\sin\phi_v,\cos\theta_v)$ is {a unit vector}. The potential breaks the rotation symmetry of the system explicitly, and the system has  minimal total energy when the vortex center coincides with $\bm{r}_v$.

In the following, we calculate the expectation of the angular momentum, which will be employed in the study of the dynamics of the vortex. For the special case with the vortex center locating at the north pole, the expectation reads
\begin{equation}
    \bm{l}=\langle\psi_{+}(0,0)|\hat{\bm{L}}|\psi_{+}(0,0)\rangle,
\end{equation}
where the angular momentum operator in the presence of a monopole is
\begin{eqnarray}
    \hat{\bm{L}}&=&\bm{r}\times(-i\nabla-\bm{A}_{+})-I\bm{r}/r \notag\\
    &=& -i\bm{e}_{\phi}\frac{\partial}{\partial\theta}+i\bm{e}_{\theta}\frac{\partial}{\sin\theta\partial\phi}+\bm{e}_{\theta}\frac{1-\cos\theta}{2\sin\theta}-\frac{\bm{e}_{r}}{2}.
\end{eqnarray}
Using the wave function \eqref{psig}, we have
\begin{equation}
\bm{l}=\int\sin\theta d\theta d\phi f^*(\theta)\Big[-i\bm{e}_{\phi}\frac{\partial}{\partial\theta}-\bm{e}_{\theta}\frac{1+\cos\theta}{2\sin\theta}-\bm{e}_{r}\frac{1}{2}\Big]f(\theta).
\end{equation}
Now we consider the three components $l_i = \bm{e}_{i}\cdot \bm l$ for $i=x,y,z$.
Note that $\bm{e}_{x}\cdot\bm{e}_{\phi}= -\sin\phi$, $\bm{e}_{x}\cdot\bm{e}_{\theta}=\cos\theta \cos\phi$,
and $\bm{e}_{x}\cdot\bm{e}_{r}=\sin\theta\cos\phi$. Therefore, $l_{x}$ vanishes since
the integral over $\phi$ is null. For the $y$ component, we have $\bm{e}_{y}\cdot\bm{e}_{\phi} = \cos\phi$, $\bm{e}_{y}\cdot\bm{e}_{\theta}=\cos\theta \sin\phi$, and $\bm{e}_{y}\cdot\bm{e}_{r}=\sin\theta\sin\phi$. Similarly, we obtain $l_{y}=0$. For the $z$ component, we have $\bm{e}_{z}\cdot\bm{e}_{\phi}=0$, $\bm{e}_{z}\cdot\bm{e}_{\theta}=-\sin\theta$
, $\bm{e}_{z}\cdot\bm{e}_{r}=\cos\theta$. As a result,
\begin{equation}
   l_{z}=\int\sin\theta d\theta d\phi f^*(\theta)\Big[\frac{1+\cos\theta}{2}-\frac{\cos\theta}{2}\Big]f(\theta)=\frac{1}{2}.
\end{equation}
Therefore, $\bm l = (0,0,1/2)$, which is the angular momentum for each atom. It is also easy to check that $|\psi_{+}(0,0)\rangle$ is indeed the eigenstate of $\hat L_z$ with eigenvalue $1/2$. We can generalize the expectation value of the angular momentum to the case with an arbitrary vortex position $(\theta_v,\phi_v)$ by a rotation transformation \eqref{sol}, which gives ${\bm l} \equiv \langle \hat{\bm L} \rangle = \bm n_v/2$. Note that the angular momentum is interaction-independent and its direction coincides with the vortex position vector.

\section{Single vortex-monopole dynamics }\label{sec5}
In this section, we study the dynamics of the single vortex induced by the minimal monopole in an additional uniform magnetic field. Without loss of generality, we suppose that the magnetic field is along the $z$ direction, ${\bm B}_2 = B_2{\bm e}_z$, and the corresponding vector potential is $\delta \bm A = B_2 (-y/2, x/2, 0)$. We assume that the monopole-vortex configuration is rigid. {This is inspired by the Thiele equation for topological particles  \cite{Thiele1973}, such as magnetic skyrmions or Hopfions  \cite{Reichhardt2022,Wang2019}.} These particles are typically treated as undeformed, particle-like entities that maintain their structure during motion, thereby simplifying their dynamics. Our assumption is also motivated by the analogy to vortex dynamics under strong interactions.  {When the healing length is small, a point vortex model is commonly employed to describe vortex motion \cite{Turner2010}, as phonon excitations are strongly suppressed in this regime.} Our theoretical predictions, derived under this assumption, show excellent agreement with numerical results, thereby confirming its validity in the strong interaction region. For more general cases, where phonon excitations may deform the vortex profile and influence its trajectory, further research will be conducted to explore these effects.

Using this assumption, the evolving wavefunction becomes $| \psi(t)\rangle \propto |\psi_{\pm}[\theta_v(t),\phi_v(t)]\rangle $ with a varying vortex center. With this ansatz, the system is described by the dynamics of the vortex center, or equivalently, the dynamics of the angular momentum ${\bm l}(t) = {\bm n}_v(t)/2$. From the GPE \eqref{gpe} and the definition $\bm l = \langle \psi(t)| \hat{\bm L} |\psi(t)\rangle$, we obtain
\begin{equation}
\dot{\bm{l}}=-i\langle\psi(t)|[\hat{\bm{L}},\hat{H}_{\text{eff}}]|\psi(t)\rangle,
\end{equation}
where the effective Hamiltonian
\begin{equation}
\hat{H}_{\text{eff}}(\delta\bm{A}) = \frac{(\hat{\bm{p}}-\delta\bm{A})^{2}}{2m}+V(r)+\lambda_{3D}|\psi|^{2}
\end{equation}
contains the gauge fields from both the monopole and the external magnetic field.
We can split the effective Hamiltonian into two parts,
\begin{eqnarray}
    \hat{H}_{\text{eff}}(\delta\bm{A}) = \hat{H}_{\text{eff}}(\delta\bm{A}=0)+\Delta\hat{H},
\end{eqnarray}
where
\begin{equation}
\Delta\hat{H}=\frac{(i\nabla\cdot\delta\bm{A})-2\delta\bm{A}\cdot\hat{\bm{p}}+\delta\bm{A}\cdot\delta\bm{A}}{2m}
\end{equation}
is the correction contributed by the external field. Using the fact that the instantaneous $|\psi(t)\rangle$ is the eigen-wavefunction of $\hat{H}_{\text{eff}}(\delta\bm{A}=0)$, we have
\begin{equation}
    \dot{\bm{l}} = -i\langle\psi(t)|[\hat{\bm{L}},\Delta\hat{H}]|\psi(t)\rangle.
\end{equation}
Note that
$\nabla\cdot\delta\bm{A}=0$, $2\delta\bm{A}\cdot\hat{\bm{p}}=B_{2}[\hat{L}_{z}+(1/2)\cos\theta]$,
and $\delta\bm{A}\cdot\delta\bm{A}=(1/4)B_{2}^{2}r^{2}\sin^{2}\theta$, the Hamiltonian becomes
\begin{equation}
    \Delta\hat{H} = -\alpha\hat{L}_{z}+g(r,\theta),
\end{equation}
where $\alpha=B_{2}/(2m)$ is the Larmor precession frequency and the scalar potential is
\begin{equation} \label{potss}
    g(r, \theta)=-\frac{B_{2}\cos\theta}{4m}+\frac{B_{2}^{2}r^{2}\sin^{2}\theta}{8m}.
\end{equation}

{Detailed} calculations give
\begin{equation}
-i\langle\psi(t)|[\hat{\bm{L}},-\alpha\hat{L}_{z}]|\psi(t)\rangle = -\alpha l_{x}\bm{e}_{y}+\alpha l_{y}\bm{e}_{x}=\bm{\omega}\times\bm{l},
\end{equation}
where $\bm{\omega}=-\alpha\bm{e}_{z}$, and
\begin{eqnarray}
&& -i \langle\psi(t)| [\hat{\bm{L}},  g(r,\theta)]|\psi(t)\rangle \notag \\
&& ~~~~~~~= -i\langle[\bm{r}\times(-i\nabla-\bm{A}_{+})-I\bm{r}/r,g(r,\theta)]\rangle \notag \\
&& ~~~~~~~=-\langle[\bm{r}\times\nabla,g(r,\theta)]\rangle=-\langle\bm{e}_{\phi}\partial_\theta g(r, \theta)\rangle.
\end{eqnarray}
 If the corresponding density of the wave function $|\psi(t) \rangle$ is exactly uniform
on the whole sphere, then we have $\langle\bm{e}_{\phi} \partial_\theta g(r, \theta) \rangle=0$ because $\bm{e}_{\phi} $ changes its direction along the circle of latitude. However, since the system has a small size of vacancy at the vortex center, and considering the shell-shaped case, $r \simeq r_0$, we have
\begin{equation}
\langle\bm{e}_{\phi} \partial_\theta g(r, \theta) \rangle=-\bm{e}_{\phi_{v}} \partial_\theta g(r_0, \theta)|_{\theta=\theta_{v}}  n_{\text{miss}}.
\end{equation}
Therefore,
\begin{equation}
\dot{\bm{l}}=\bm{\omega}\times\bm{l}+\bm{\tau},
\end{equation}
where $\bm{\tau}= \partial_\theta g(r_0, \theta)|_{\theta=\theta_{v}} n_{\text{miss}}\bm{e}_{\phi_{v}}$ is a torque contributed by the potential $g(r,\theta)$. In the following, we provide an interpretation of this result from a classical perspective. The effective energy for a vortex in the potential $g(r_0, \theta)$ is $E(\theta_v) = -g(r_0, \theta_v) n_{miss}$ due to the particle missing,
then, the torque is $\bm \tau  = \nabla_{\bm l} E(\theta_v) \times \bm l$, which is exactly the same as the result given above. As a result, $l_z$ is conserved during the evolution. Therefore, $\theta_v$ remains a constant value, while the precession angular frequency is
\begin{equation}\label{omega}
 \omega_{m} \equiv \dot{\phi_{v}}(t)= -\alpha + \frac{\partial_\theta g(r_0, \theta)|_{\theta=\theta_{v}} n_{\text{miss}}}{|\bm l|\sin\theta_{v}}.
\end{equation}
where $\omega_m = - (1- n_{\text{miss}}) {B_2}/{2m} +n_{\text{miss}}{B_{2}^{2}r_{0}^{2}\cos\theta_{v}}/{2m}$
arise from the magnetic field $B_2 \hat{\bm e}_z$. The external magnetic field drives the precession of the vortex. $\omega_m$ is slightly different from the Larmor precession frequency due to the particle missing in the vortex. When considering the influence of a microgravity or an electric field along the $z$ direction, which gives an additional contribution in the scalar potential ($\propto \cos\theta$) in \eqref{potss}, the precession frequency will be further corrected.

\section{Numerical simulations}\label{sec4}
In this section, we numerically simulate the vortex dynamics in the presence of $\bm B_2$ and compare it with the analytical results. For this purpose, we first develop the numerical approach. To avoid the singularities (Dirac strings) of the gauge field, we use the bi-gauge-evolution method to cover the whole sphere.

\subsection{Bi-gauge evolution of the GPE}
The dynamics of BEC in the presence of a monopole are described by GPE \eqref{gpe},
where the regions for $\psi_{+}$ and $\psi_{-}$ are $R_{\pm}$, respectively. The two wave functions in the overlapped region are linked by the gauge transformation $\psi_{-}(\textbf{r})=S^{-1}\psi_{+}(\textbf{r})$. We use the split-step Crank-Nicolson algorithm to solve the GPE. This approach has been introduced in details in Refs.\,\cite{Adhikari2009,Kumar2019}. In the current system, we split the Hamiltonian into four parts: $H_{ \pm}=H_{1}^{\pm}+H_{2}^{\pm}+H_{3}^{\pm}+H_{4}^{\pm}$, where
\begin{eqnarray}
 H_{1}^{\pm} &=& V+\frac{\left|\bm A_{ \pm}\right|^2}{2m}+\lambda_{3D}\left|\psi_{ \pm}\right|^2, \\
 H_{2}^{\pm} &=& -\frac{1}{2m} \frac{\partial}{\partial x^2} + \frac{i}{2} A_{x}^{ \pm} \frac{\partial}{\partial x} + \frac{i}{2} \frac{\partial}{\partial x} A_{x}^{ \pm}, \\
 H_{3}^{\pm} &=& -\frac{1}{2m} \frac{\partial}{\partial y^2} + \frac{i}{2m} A_{y}^{ \pm} \frac{\partial}{\partial y} + \frac{i}{2m} \frac{\partial}{\partial y} A_{y}^{ \pm}, \\
 H_{4}^{\pm} &=& -\frac{1}{2m} \frac{\partial}{\partial z^2} + \frac{i}{2m} A_{z}^{ \pm} \frac{\partial}{\partial z} + \frac{i}{2m} \frac{\partial}{\partial z} A_{z}^{ \pm}.
\end{eqnarray}
In the 3D numerical simulations, we use $150\times 150 \times 150$ grids for space and discretize time into $\{\cdots, t_{n-1}, t_n, \cdots\}$ with spacing $\Delta t$. For each gauge choice, the evolution operator is split into four steps:
\begin{equation}\label{x}
  e^{-i H_{ \pm} \Delta t} = e^{-i H_4^{ \pm} \Delta t} e^{-i H_3^{ \pm} \Delta t} e^{-i H_2^{ \pm} \Delta t} e^{-i H_1^{ \pm} \Delta t},
\end{equation}
and $\underline{\psi}_\pm^{n+1} = e^{-i H_{ \pm} \Delta t} {\psi}_\pm^{n}$.
More details can be seen in Appendix \ref{cna}.

\begin{figure}[tbp]
\includegraphics[angle=0,width=0.95\columnwidth]{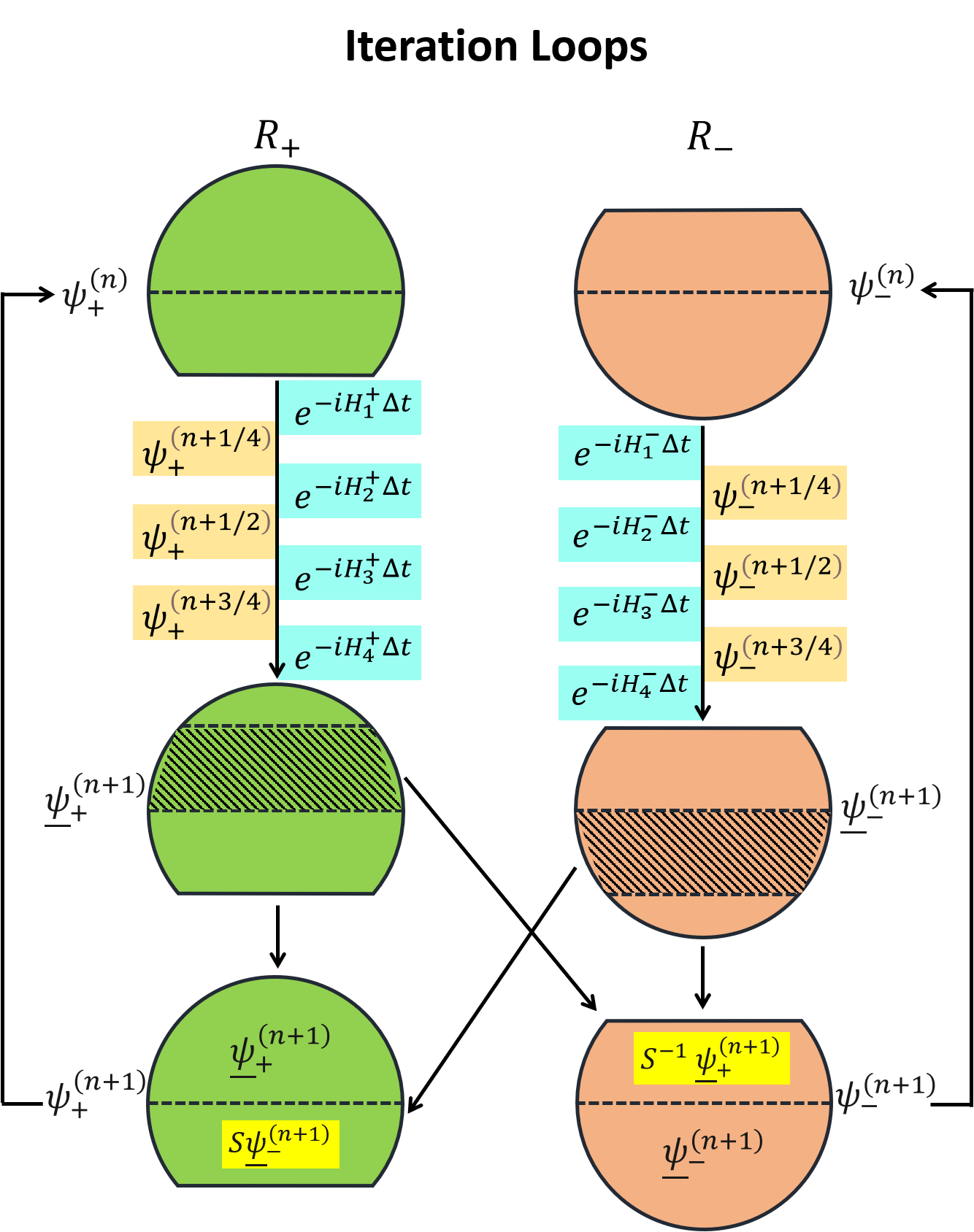}
\caption{Schematic illustration of the iteration }
\label{sm1}
\end{figure}

To avoid Dirac strings, we use two gauges in the two regions: $R_+$ and $R_-$. Each region has an open boundary. The boundary would induce non-physical effects on the profile of the wavefunction in numerical simulations. In order to get the correct behavior of the boson gas on the shell with a monopole, we bridge the two wave functions at each iteration:
\begin{equation}
    {\psi}_+^{n+1} = \left\{
    \begin{array}{ccc}
         \underline{\psi}_+^{n+1}  &  & \text{for}~~ z \geq 0 \\
        S \underline{\psi}_-^{n+1}  &  &  \text{for}~~ z < 0,
    \end{array}
    \right.
\end{equation}
and
\begin{equation}
{\psi}_-^{n+1} =  \left\{
    \begin{array}{ccc}
     S^{-1} \underline{\psi}_+^{n+1}  &  &  \text{for}~~   z > 0 \\
       \underline{\psi}_-^{n+1}   &  &  \text{for}~~  z \leq 0.
    \end{array}
    \right.
\end{equation}
These finish one iteration. We sketch the iteration loop in Fig.\,\ref{sm1}. The bi-gauge evolution eliminates the non-physical effects from boundary and overcomes the singularity in the one-gauge method, so we can get the correct wavefunction.

In the imaginary-time evolution simulation, we rigorously monitor the convergence of both the chemical potential and the wavefunction.  {We successfully reproduce the stable vortex lattice configuration in a spherical BEC with integer-type magnetic monopoles \cite{Zhou2018}, thereby validating the numerical method}.

\begin{figure*}[t]
\includegraphics[angle=0,width= \textwidth]{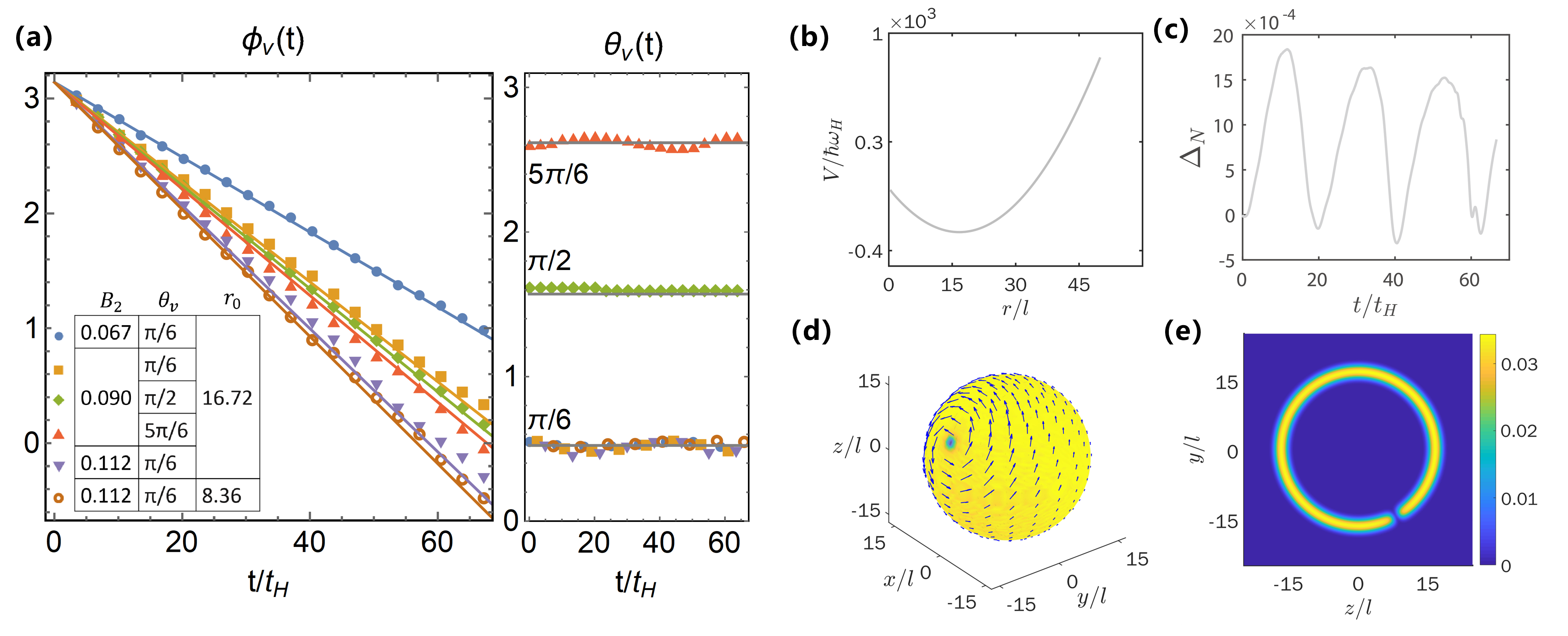}
\caption{(a) The single-vortex core trajectory under a magnetic field: $\phi_v(t)$ (left) and $\theta_v(t)$ (right). All lines are the analytical results \eqref{omega} of the evolving vortex center, and all symbols are 3D numerical simulation results for different magnetic field strengths $B_2$ (unit $[m\omega_H]$), initial vortex core positions $\theta_v$, and radii $r_0$ (unit $\ell\equiv \sqrt{\hbar/m\omega_H}$). Here, $\lambda_{3D} = 13375 ~\hbar\omega_H \ell^3$, $t_H = 1/\omega_H$, and the initial $\phi_v$ equals to $\pi$. (b) The scalar shell potential $V(r)$ for the case of $r_0 =16.72 \ell$. (c) The deviation in the normalization of the wavefunction during dynamical evolution,   $\Delta_N = \int d\bm r |\psi(\bm r)|^2 -1$. (d) The velocity distribution and the density profile of the BEC on the shell with a single vortex at  $(\theta_v = \pi/3, \phi_v =3\pi/2)$ and (e) the corresponding density profile on the $x=0$ plane.}
\label{fig2t}
\end{figure*}

\subsection{Simulation results}

In this subsection, we present our 3D numerical simulations of vortex trajectory on a narrow shell. For strong interaction with a thin shell, we have bridged the 3D interaction strength and the effective 2D interaction strength in Appendix \ref{inter}, by using the Thomas-Fermi approximation along the radial direction. Consequently, we can compare the 3D numerical simulation with the 2D analytical results shown in Sec.\ref{sec5}. The initial state in the simulation is prepared by the imaginary-time evolution of the GPE \cite{Adhikari2009,Kumar2019}. The initial position of the vortex can be controlled by adding a small positive Gaussian potential to pin it, as illustrated in Sec. \ref{sec3}. The real-time dynamical simulation is implemented after switching on $\delta \bm A$. To prevent singularities in the simulation, we utilize the bi-gauge numerical method that we developed in the previous subsection.

In Fig.\,\ref{fig2t}(a), we present the numerical results of the vortex trajectory. From the result of $\theta_v(t)$, we observe that $\theta_v(t)$ almost remains constant during the volution, while $\phi_v(t)$ changes linearly with time. This shows that the vortex experiences a precession around the $z$-axis.

With a fixed sphere radius of $r_0 = 16.72 \ell$ (length unit $\ell = \sqrt{\hbar/m\omega_H}$), we further confirm that as the external magnetic field $B_2$ gets stronger, the vortex's precession frequency gets stronger. For a vortex initially posed at $\theta_v = \pi/6$, the simulated results agree very well with our prediction for different magnitudes of ${B_2} = 0.067$, $0.09$, and $0.112$ (unit $m\omega_H$).

To see how the initial position of the vortex influences the precession, for fixed ${B_2} = 0.09~m\omega_H$, we select different initial vortex positions at $\theta_v =\pi/6$, $\pi/2$, and $5\pi/6$, respectively. The value of initial $\theta_v$ slightly changes the precession frequency, and the results qualitatively agree with our analytic prediction.

We also try to check the effect of the shell radius. For ${B_2} = 0.112~m\omega_H$, we simulate the motion for the vortex (with initial position $\theta_v =\pi/6$) at the shell with radius $r_0 = 16.72 \ell$ and $ 8.36 \ell$, respectively. The results again qualitatively agree with our prediction in \eqref{omega}.

{In Fig.\,\ref{fig2t}(b-e), we present additional details of the simulation. Fig.~\ref{fig2t}(b) shows the scalar potential $V(r)$ in Eq.\,\eqref{eq12} for the case of $r_0 = 16.72\ell$. In the real-time evolution, we monitor the normalization of the wavefunction, which should remain constant throughout the evolution. Fig.\,\ref{fig2t}(c) illustrates the deviation of the normalization during dynamical evolution, defined as  $\Delta_N = \int d\bm r |\psi(\bm r)|^2 -1$. We confirm that the deviation is kept below 0.2\%.  Fig.\,\ref{fig2t}(d) visualizes the velocity distribution and the density profile of the BEC on the shell with a single vortex at $(\theta_v = \pi/3, \phi_v =3\pi/2)$, while Fig.\,\ref{fig2t}(e) shows the corresponding density profile on the \(x=0\) plane.}

\section{Conclusion}

In conclusion, we have developed a pioneering method that enables the realization of magnetic monopoles with half-integer charges in the BEC systems. We discussed the suggested parameters for the experiment setup and explored the experimental flexibility. We have systemically studied ground-state vortex solutions and single-vortex dynamics on a sphere, which includes analytical calculations under the assumption of a rigid vortex-monopole structure and numerical simulations using a novel method to prevent singularities of the gauge fields. Larmor precession with a modified frequency captures the vortex dynamics on the sphere in the presence of a uniform electromagnetic field.

We note that the spherical BEC itself has garnered intense attention due to its nontrivial geometric properties. The bubble system realized in microgravity environments is also a platform to simulate cosmic inflation \cite{Tononi2019,Carollo2022,Tononi2020,Brooks2021,He2022,Rhyno2021,Jia2022}. The nontrivial topology and the finite size endow phase transitions in spherical systems with novel characters \cite{Tononi2019,Tononi2022}. On a sphere, the vortex and antivortex always appear in pairs when there are no monopoles \cite{Bogomolov1977}. The existence of monopoles enables the investigation of vortex dynamics in a wider range of configurations.

Our research substantially advances the realization of magnetic monopoles, enhances our understanding of vortex dynamics and monopole properties, lays the foundation for experimental verification, and is helpful for further study of vortex dynamics in closed manifolds.

\begin{acknowledgments}
We acknowledge the discussion from Congjun Wu, Lichen Zhao, Shanchao Zhang, and Biao Wu. This work is supported by the National Natural Science Foundation of China under grants No.\,12105223, No.\,12175180, No.\,11934015, No.\,12305029, and No.\,12247103, the Major Basic Research Program of Natural Science of Shaanxi Province under grants No.\,2017KCT-12 and No.\,2017ZDJC-32,  Shaanxi Fundamental Science Research Project for Mathematics and Physics under grant No.\,22JSZ005 and No.\,22JSQ041, and Natural Science Basic Research Program of Shaanxi under Grant No.\,2024JC-YBMS-022 and No.\,2023-JC-QN-0054. This research is also supported by the Youth Innovation Team of Shannxi Universities and The Double First-class University Construction Project of Northwest University.
\end{acknowledgments}

\appendix{}

\section{2D effective interaction}\label{inter}
Here, we derive the 2D effective interaction strength $\lambda_{2D}$ for a shell-shaped BEC with large $r_0$. The Gross-Pitaevskii equation (GPE) is
\begin{eqnarray}\label{gpe2ap}
    i\hbar{\partial_t{\psi_{\pm}}} =  \frac{(-i\nabla -
    \bm{A}_{\pm})^{2}}{2m} \psi_{\pm} + V \psi_{\pm} +\lambda_{3D}\left|\psi_{\pm}\right|^{2}\psi_{\pm},
\end{eqnarray}
where ${V}(r) = {V}_H(r) - 2\mu_I B'_{1}r + I/2m r^2$. For a large shell geometry, we approximately have ${V}(r) = \frac{1}{2} m \omega_H^2 (r-r_0)^2$, after discarding a constant. We write the wave function as $\psi = \frac{1}{r} g(r) f(\theta, \phi, t)$. The normalization condition is
\begin{eqnarray}
    1 &=& \int g^2(r) dr, \\
    1 &=& \int\int |f(\theta, \phi, t)|^2 \sin\theta d \theta d\phi.
\end{eqnarray}
To get $\lambda_{2D}$, {We left-multiply the GPE \eqref{gpe2ap} by $r g(r)$ and then perform a radial integration $\int dr$}. Then, we obtain
\begin{equation}\label{gpe21}
    i\hbar{\partial_t{f}}(\theta,\phi,t) = \cdots +\lambda_{3D}\left|f\right|^{2}f \int \frac{1}{r^2} g^4(r) dr.
\end{equation}
We denote
\begin{equation}\label{gpe22}
    \frac{\lambda_{2D}}{r_0^2} \equiv \lambda_{3D} \int \frac{1}{r^2} g^4(r) dr.
\end{equation}
For $r_0 \gg \ell$, we have
\begin{equation}\label{gpe2x}
    {\lambda_{2D}}\simeq \lambda_{3D} \int g^4(r) dr.
\end{equation}
In the static case,  we use the Thomas-Fermi approximation,
\begin{equation}\label{gpe23}
   \mu_F {\psi} \simeq V \psi +\lambda_{3D}\left|\psi_{\pm}\right|^{2}\psi.
\end{equation}
Assuming that the density is uniform for different $\theta$ and $\phi$, i.e., $|f|^2 \simeq 1/4\pi$, we obtain
\begin{equation}\label{gpe24}
   \mu_F =  \frac{1}{2} m \omega_H^2 (r-r_0)^2 +\lambda_{3D} g^2(r)/ 4\pi r^2.
\end{equation}
Thus, for large $r_0$,
\begin{eqnarray}
    g^2(r) &=& 4 \pi r^2 [\mu_F- m \omega_H^2 (r-r_0)^2 /2 ]/ \lambda_{3D} \notag \\
     &\simeq& 4\pi r_0^2 [\mu_F- m \omega_H^2 (r-r_0)^2 /2 ]/ \lambda_{3D}
\end{eqnarray}
We denote $\mu_F = m \omega_H^2 R^2_{TF} /2$. Then from the normalization of $g(r)$, we have
\begin{equation}
  1= \frac{8\pi}{3} \frac{m \omega_H^2 r_0^2 R^3_{TF} }{ \lambda_{3D}}.
\end{equation}
Finally, we obtain
\begin{eqnarray}\label{gpe25}
    {\lambda_{2D}} &=& \lambda_{3D} \int g^4(r) dr \notag\\
    &=&\frac{2}{5}(9 \pi m \omega_H^2 r_0^2 \lambda_{3D}^2) ^{ \frac{1}{3}}.
\end{eqnarray}

\section{Crank-Nicolson algorithm for gauge field GPE}\label{cna}
Here, we illustrate the discretization of GPE. When the time spacing $\Delta t$ is sufficient small, for each gauge choice, the evolution operator is split into four steps:
\begin{equation}\label{x7}
  e^{-i H_{ \pm} \Delta t} = e^{-i H_4^{ \pm} \Delta t} e^{-i H_3^{ \pm} \Delta t} e^{-i H_2^{ \pm} \Delta t} e^{-i H_1^{ \pm} \Delta t}.
\end{equation}
we solve the GPE step by step individually:

(1) In the first step, since $H_{1}^{ \pm}$ is already diagonalized,
\begin{equation}
\psi_\pm^{n+1/4}=e^{-i H^{\pm}_{1} \Delta t }\psi_\pm^{n}
\end{equation}
can be implemented directly.

(2) In the second step,
$\psi_\pm^{n+1/2}=e^{-i H^{\pm}_{2} \Delta t }\psi_\pm^{n+1/4}$.
Linear expansion gives ($m =1$)
\begin{equation}\label{eomstep2}
\begin{aligned}
& i \frac{\psi_{\pm, j}^{n+1 / 2}-\psi_{\pm, j}^{n+1 / 4}}{\Delta t}=-\frac{1}{2} \frac{1}{2 \Delta_x^2}\Big\{\big(\psi_{\pm, j+1}^{n+1 / 2}-2 \psi_{\pm, j}^{n+1 / 2} \\
& +\psi_{\pm, j-1}^{n+1 / 2}\big)+\big(\psi_{\pm, j+1}^{n+1/4}-2 \psi_{\pm j}^{n+1/4}+\psi_{\pm j-1}^{n+1/4}\big)\Big\} \\
& +\frac{i A_{x, j}^\pm}{8 \Delta_x}\Big\{\big(\psi_{\pm, j+1}^{n+1 / 2}-\psi_{\pm, j-1}^{n+1 / 2}\big)+\big(\psi_{\pm,j+1}^{n+1/4}-\psi_{\pm,j-1}^{n+1/4}\big)\Big\} \\
& +\frac{i}{8 \Delta_x}\Big\{\big(A_{x,j+1}^{\pm}\psi_{\pm,j+1}^{n+1 / 2}-A_{x,j-1}^{\pm}\psi_{\pm,j-1}^{n+1/2}\big) \\
& +\big(A_{x,j+1}^{\pm}\psi_{\pm,j+1}^{n+1 / 4}-A_{x,j-1}^{\pm}\psi_{\pm,j-1}^{n+1/4}\big)\Big\},         \end{aligned}
\end{equation}
where $j=1,2,3,...$ is the $j$-th grid point in $x$ direction for given $y$ and $z$, and $\Delta_x$ is the spacing. Eq.\,\eqref{eomstep2} results in a set of equations
\begin{equation}\label{x}
  \alpha_{\pm, j}^{-} \psi_{\pm, j-1}^{n+1 / 2}+\alpha_{\pm, j}^0 \psi_{\pm, j}^{n+1 / 2}+\alpha_{\pm, j}^{+} \psi_{\pm, j+1}^{n+1 / 2}=\beta_{\pm, j},
\end{equation}
where
\begin{equation}\label{eq39}
\begin{aligned}
\beta_{\pm,j}  = & \frac{i \Delta t}{4 \Delta_x^2}\big(\psi_{\pm, j+1}^{n+1 / 4}-2 \psi_{\pm,j}^{n+1 / 4}+\psi_{\pm,j-1}^{n+1 / 4}\big) \\
& +\frac{\Delta t}{8 \Delta_x}\Big\{ A_{x, j}^\pm\big(\psi_{\pm, j+1}^{n+1 / 4}-\psi_{\pm, j-1}^{n+1 / 4}\big) \\
&+\big(A_{x,j+1}^{\pm}\psi_{\pm,j+1}^{n+1 / 4}-A_{x,j-1}^{\pm}\psi_{\pm,j-1}^{n+1 /4}\big) \Big\} \\
\alpha_{\pm,j}^0  = & 1+\frac{i \Delta t}{2 \Delta_x^2}, \\
 \alpha_{\pm, j}^{-} =&-\frac{i \Delta t }{8 \Delta_x}\Big[\frac{2}{\Delta_x}+ i (A_{x,j-1}^{\pm}+A_{x,j}^\pm)\Big], \\
\alpha_{\pm,j}^{+}  = & -\frac{i \Delta t}{8 \Delta_x}\Big[\frac{2}{\Delta_x}-i (A_{x,j +1}^{\pm}+A_{x,j}^\pm)\Big].
\end{aligned}
\end{equation}
Then, the wave function $\psi_\pm^{n+1/2}$ is obtained accordingly. A similar method can be performed for the third step $
\psi_\pm^{n+3/4} = e^{-i H_3^\pm \Delta t} \psi_\pm^{n+1/2}$ and the fourth step $
  \underline{\psi}_\pm^{n+1} = e^{-i H_4^\pm \Delta t}\psi_\pm^{n+3/4}$.

\newpage


\begin{thebibliography}{1}
\bibitem{Dirac1931}
P. A. M. Dirac, Quantised singularities in the electromagnetic field, \href{https://doi.org/10.1098/rspa.1931.0130}{Proc. R. Soc. A \textbf{133}, 60 (1931)}.

\bibitem{Yang1975}
T. T. Wu and C. N. Yang, Concept of
nonintegrable phase factors and global formulation of gauge fields, \href{http://dx.doi.org/10.1103/PhysRevD.12.3845}{Phys. Rev. D \textbf{12}, 3845 (1975)}.

\bibitem{Yang1976}
T. T. Wu and C. N. Yang, Dirac monopole
without strings: Monopole harmonics, \href{https://doi.org/10.1016/0550-3213(76)90143-7}{Nucl. Phys. B \textbf{107}, 365 (1976)}.

\bibitem{Drukier1979}
M. Bonnardeau and A. K. Drukier, Creation of magnetic monopoles in pulsars, \href{https://doi.org/10.1038/277543a0}{Nature \textbf{277}, 543 (1979)}.

\bibitem{Friseh1990}
H. J. Friseh, Quest for magnetic monopoles, \href{https://doi.org/10.1038/344706a0}{Nature \textbf{344}, 706 (1990)}.

\bibitem{Shellard2000}
A. Vilenkin and E. P. S. Shellard, {\it Cosmic Strings and Other Topological Defects}, Cambridge Monographs on Mathematical Physics (Cambridge University Press, Cambridge, 2000).

\bibitem{Guth1981}
A. H. Guth, Inflationary universe: A possible solution to the horizon and flatness problems, \href{http://dx.doi.org/10.1103/PhysRevD.23.347}{Phys. Rev. D \textbf{23}, 347 (1981)}.


\bibitem{Salomaa1987}
M. M. Salomaa, Monopoles in the rotating superfluid Helium-3 a-b interface, \href{http://dx.doi.org/10.1038/326367a0}{Nature \textbf{326}, 367 (1987)}.

\bibitem{Castelnovo2008}
C. Castelnovo, R. Moessner, and S. L. Sondhi, Magnetic monopoles in spin ice, \href{http://dx.doi.org/10.1038/nature06433}{Nature \textbf{451}, 42 (2008)}.

\bibitem{Gingras2009}
M. J. P. Gingras, Observing monopoles in a
magnetic analog of ice, \href{http://dx.doi.org/10.1126/science.1181510}{Science \textbf{326}, 375 (2009)}.

\bibitem{Bramwell2009}
S. T. Bramwell, S. R. Giblin, S. Calder, R. Aldus, D. Prabhakaran, and T. Fennell, Measurement of the charge and current of magnetic monopoles in spin ice,
\href{http://dx.doi.org/10.1038/nature08500}{Nature \textbf{461}, 956 (2009)}.

\bibitem{Giblin2011}
S. R. Giblin, S. T. Bramwell, P. C. W. Holdsworth, D. Prabhakaran, and I. Terry, Creation and measurement of long-lived magnetic monopole currents in spin ice, \href{http://dx.doi.org/10.1038/nphys1896}{Nat. Phys. \textbf{7}, 252 (2011)}.

\bibitem{Lin2016}
S.-Z. Lin and A. Saxena, Dynamics of Dirac
strings and monopole like excitations in chiral magnets under a current drive, \href{http://dx.doi.org/10.1103/PhysRevB.93.060401}{Phys. Rev. B \textbf{93}, 060401 (2016)}.

\bibitem{Fang2003}
Z. Fang, N. Nagaosa, K. S. Takahashi, A. Asamitsu, R. Mathieu, T. Ogasawara, H. Yamada, M. Kawasaki, Y. Tokura, and K. Terakura, The anomalous Hall effect and magnetic monopoles in momentum space, \href{http://dx.doi.org/10.1126/science.1089408}{Science \textbf{302}, 92 (2003)}.

\bibitem{Boxem2013}
A. B\'ech\'e, R. V. Boxem, G. V. Tendeloo, and J. Verbeeck, Magnetic monopole field exposed by electrons, \href{https://doi.org/10.1038/nphys2816}{Nature
Physics \textbf{10}, 26 (2013)}.

\bibitem{Zhang2017}
Z.-L. Zhang, M.-F. Chen, H.-Z. Wu, and Z.-B. Yang, Quantum simulation of abelian Wu-Yang monopoles in spin-1/2 systems, \href{http://dx.doi.org/10.1088/1612-202X/aa622b}{Laser Phys. Lett. \textbf{14}, 045205 (2017)}.

\bibitem{Edelstein2024}
M. I. Marqu\'es, S. Edelstein, P. A. Serena, B. C. L. de Larrinzar, and A. Garcia-Mart\'in,
Magneto-optical particles in isotropic spinning fields mimic magnetic monopoles, \href{https://doi.org/10.1103/PhysRevLett.133.046901}{Phys. Rev. Lett. \textbf{133}, 046901 (2024)}.

\bibitem{Ray2014}
M. W. Ray, E. Ruokokoski, S. Kandel, M. M\"ott\"onen, and D. S. Hall, Observation of Dirac monopoles in a synthetic magnetic field, \href{http://dx.doi.org/10.1038/nature12954}{Nature \textbf{505}, 657 (2014)}.

\bibitem{Pietil2009}
V. Pietil\"a and M. M\"ott\"onen, Creation of Dirac monopoles in spinor Bose-Einstein condensates, \href{http://dx.doi.org/10.1103/PhysRevLett.103.030401}{Phys. Rev. Lett. \textbf{103}, 030401 (2009)}.

\bibitem{Ray2015}
M. W. Ray, E. Ruokokoski, K. Tiurev, M. M\"ott\"onen, and D. S. Hall, Observation of isolated monopoles in a quantum field, \href{https://www.science.org/doi/10.1126/science.1258289}{Science \textbf{348}, 544 (2015)}.

\bibitem{Ruokokoski2011}
E. Ruokokoski, V. Pietil\"a, and M. M\"ott\"onen, Ground-state Dirac monopole, \href{http://dx.doi.org/10.1103/PhysRevA.84.063627}{Phys. Rev. A \textbf{84}, 063627 (2011)}.




\bibitem{Xu2024}
G.-S. Xu, M. Jain, X.-F. Zhou, G.-C. Guo, M. A. Amin, H. Pu, and Z.-W. Zhou, Engineering and revealing Dirac strings in spinor condensates, \href{https://doi.org/10.1103/PhysRevResearch.6.023272}{Phys. Rev. Research \textbf{6}, 023272 (2024)}.

\bibitem{Zhou2018}
X.-F. Zhou, C. Wu, G.-C. Guo, R. Wang, H. Pu, and Z.-W. Zhou, Synthetic Landau levels and spinor vortex matter on a Haldane spherical
surface with a magnetic monopole, \href{http://dx.doi.org/10.1103/PhysRevLett.120.130402}{Phys. Rev. Lett. \textbf{120}, 130402 (2018)}.




\bibitem{Turner2010}
A. M. Turner, V. Vitelli, and D. R. Nelson,
Vortices on curved surfaces, \href{http://dx.doi.org/10.1103/RevModPhys.82.1301}{Rev. Mod. Phys. \textbf{82}, 1301 (2010)}.




\bibitem{Sun2020}
K. Padavi\'c, K. Sun, C. Lannert, and
S. Vishveshwara, Vortex-antivortex physics in
shell-shaped Bose-Einstein condensates, \href{http://dx.doi.org/10.1103/PhysRevA.102.043305}{ Phys. Rev. A \textbf{102}, 043305 (2020)}.




\bibitem{Guenther2020}
N.-E. Guenther, P. Massignan, and A. L. Fetter, Superfluid vortex dynamics on a torus and other toroidal surfaces of revolution, \href{http://dx.doi.org/10.1103/PhysRevA.101.053606}{Phys. Rev. A \textbf{101}, 053606 (2020)}.


\bibitem{Bereta2021}
S. J. Bereta, M. A. Caracanhas, and A. L. Fetter, Superfluid vortex dynamics on a spherical film,
\href{https://doi.org/10.1103/PhysRevA.103.053306}{Phys. Rev. A \textbf{103}, 053306 (2021)}.


\bibitem{Caracanhas2022}
M. A. Caracanhas, P. Massignan, and A. L. Fetter, Superfluid vortex dynamics on an ellipsoid and other surfaces of revolution, \href{http://dx.doi.org/10.1103/PhysRevA.105.023307}{Phys. Rev. A \textbf{105}, 023307 (2022)}.



\bibitem{Tononi2022}
A. Tononi, A. Pelster, and L. Salasnich,
Topological superfluid transition in bubble-trapped condensates, \href{http://dx.doi.org/10.1103/PhysRevResearch.4.013122}{Phys. Rev. Res. \textbf{4}, 013122 (2022)}.

\bibitem{Dritschel2015}
D. G. Dritschel, M. Lucia, and Andrew C. Poje, Ergodicity and spectral cascades in point vortex flows on the sphere, \href{https://doi.org/10.1103/PhysRevE.91.063014}{Phys. Rev. E \textbf{91}, 063014 (2015)}.

\bibitem{Kanai2021}
T. Kanai and W. Guo, True mechanism of spontaneous order from turbulence in two-dimensional superfluid manifolds,
\href{https://doi.org/10.1103/PhysRevLett.127.095301}{Phys. Rev. Lett. \textbf{127}, 095301 (2021)}.

\bibitem{Yu2024}
Y. Xiong and X. Yu, Hydrodynamics of quantum vortices on a closed surface, \href{https://doi.org/10.1103/PhysRevResearch.6.013133}{Phys. Rev. Research \textbf{6}, 013133 (2024)}.

\bibitem{Efimkin2023}
G. Li and D. K. Efimkin, Equatorial waves
in rotating bubble-trapped superfluids, \href{http://dx.doi.org/10.1103/PhysRevA.107.023319}{Phys. Rev. A \textbf{107}, 023319 (2023)}.

\bibitem{White2024}
A. C. White, Triangular vortex lattices and giant vortices in rotating bubble Bose-Einstein condensates,
\href{http://dx.doi.org/10.1103/PhysRevA.109.013301}{Phys. Rev. A \textbf{109}, 013301 (2024)}.



\bibitem{He2023}
Y. He and C.-C. Chien, Vortex structure and
spectrum of an atomic fermi superfluid in a spherical bubble trap, \href{http://dx.doi.org/10.1103/PhysRevA.108.053303}{Phys. Rev. A \textbf{108}, 053303 (2023)}.


\bibitem{Haldane1983}
F. D. M. Haldane, Fractional quantization of the Hall effect: A hierarchy of incompressible quantum fluid states, \href{http://dx.doi.org/10.1103/PhysRevLett.51.605}{Phys. Rev. Lett. \textbf{51}, 605 (1983)}.



\bibitem{Pethick2008}
C. J. Pethick and H. Smith, {\it Bose-Einstein Condensation in Dilute Gases}, 2nd ed. (Cambridge University Press, Cambridge, 2008).


\bibitem{Diermaier2017}
M. Diermaier, C. B. Jepsen, B. Kolbinger, C. Malbrunot, O. Massiczek, C. Sauerzopf, M. C. Simon, J. Zmeskal, and E. Widmann, In-beam measurement of the hydrogen hyperfine splitting and prospects for antihydrogen spectroscopy, \href{http://dx.doi.org/10.1038/ncomms15749}{Nat. Commun. \textbf{8}, 15749 (2017)}.

\bibitem{Arimondo1977}
E. Arimondo, M. Inguscio, and P. Violino, Experimental determinations of the hyperfine structure in the alkali atoms, \href{http://dx.doi.org/10.1103/RevModPhys.49.31}{Rev. Mod. Phys. \textbf{49}, 31 (1977)}.




\bibitem{Thiele1973}
A. A. Thiele, Steady-State Motion of Magnetic Domains, \href{https://doi.org/10.1103/PhysRevLett.30.230}{Phys. Rev. Lett. \textbf{30}, 230 (1973)}.

\bibitem{Reichhardt2022}
C. Reichhardt, C. J. O. Reichhardt, and M. V. Milo\v{s}evi\'{c}, Statics and dynamics of skyrmions interacting with disorder
 and nanostructures, \href{https://doi.org/10.1103/RevModPhys.94.035005}{Rev. Mod. Phys. \textbf{94}, 035005 (2022)}.

\bibitem{Wang2019}
X. S. Wang, A. Qaiumzadeh, and A. Brataas, Current-Driven Dynamics of Magnetic Hopfions, \href{https://doi.org/10.1103/PhysRevLett.123.147203}{Phys. Rev. Lett. \textbf{123}, 147203 (2019)}.





\bibitem{Adhikari2009}
P. Muruganandam and S. Adhikari, Fortran programs for the time-dependent Gross-Pitaevskii equation in a fully anisotropic trap, \href{https://doi.org/10.1016/j.cpc.2009.04.015}{Comput. Phys. Commun. \textbf{180}, 1888 (2009)}.

\bibitem{Kumar2019}
R. Kishor Kumar, V. Lon\u{c}ar, P. Muruganandam, S. K. Adhikari, and A. Bala\u{z}, C and Fortran OpenMP programs for rotating Bose-Einstein condensates, \href{https://doi.org/10.1016/j.cpc.2019.03.004}{Comput. Phys. Commun. \textbf{240}, 74 (2019)}.


\bibitem{Tononi2019}
A. Tononi and L. Salasnich, Bose-Einstein condensation on the surface of a sphere, \href{http://dx.doi.org/10.1103/PhysRevLett.123.160403}{Phys. Rev. Lett. \textbf{123}, 160403 (2019)}.

\bibitem{Tononi2020}
A. Tononi, F. Cinti, and L. Salasnich, Quantum bubbles in microgravity,
\href{https://doi.org/10.1103/PhysRevLett.125.010402}{Phys. Rev. Lett. \textbf{125}, 010402 (2020)}.

\bibitem{Carollo2022}
R. A. Carollo, D. C. Aveline, B. Rhyno, S. Vishveshwara, C. Lannert, J. D. Murphree, E. R. Elliott, J. R. Williams, R. J. Thompson, and N. Lundblad, Observation of ultracold atomic bubbles in orbital microgravity, \href{http://dx.doi.org/10.1038/s41586-022-04639-8}{Nature \textbf{606}, 281 (2022)}.

\bibitem{Rhyno2021}
B. Rhyno, N. Lundblad, D. C. Aveline, C. Lannert, and S. Vishveshwara, Thermodynamics in expanding shell-shaped Bose-Einstein condensates, \href{https://doi.org/10.1103/PhysRevA.104.063310}{Phys. Rev. A \textbf{104}, 063310 (2021)}.

\bibitem{Jia2022}
F. Jia, Z. Huang, L. Qiu, R. Zhou, Y. Yan, and D. Wang, Expansion dynamics of a shell-shaped Bose-Einstein condensate, \href{https://doi.org/10.1103/PhysRevLett.129.243402}{Phys. Rev. Lett. \textbf{129}, 243402 (2022)}.

\bibitem{He2022}
Y. He, H. Guo, and C.-C. Chien, BCS-BEC crossover of atomic Fermi superfluid in a spherical bubble trap, \href{https://doi.org/10.1103/PhysRevA.105.033324}{Phys. Rev. A \textbf{105}, 033324 (2022)}.

\bibitem{Brooks2021}
M. Brooks, M. Lemeshko, D. Lundholm, and E. Yakaboylu, Molecular impurities as a realization of anyons on the two-sphere,
\href{https://doi.org/10.1103/PhysRevLett.126.015301}{Phys. Rev. Lett. \textbf{126}, 015301 (2021)}.


\bibitem{Bogomolov1977}
V. A. Bogomolov, Dynamics of vorticity at a sphere, \href{https://doi.org/10.1007/BF01090320}{Fluid Dyn. \textbf{12}, 863(1977)}.











\end{thebibliography}
\end{document}